\documentstyle[12pt,amsbsy]{article}
\newcommand{\beq}{\begin{equation}}
\newcommand{\eeq}{\end{equation}}

\newcommand{\beqn}{\begin{eqnarray}}
\newcommand{\eeqn}{\end{eqnarray}}

\newcommand{\ep}{\mbox{${\epsilon}$}}

\begin{document}

\begin{titlepage}
\vspace{1cm}

\begin{center}
{\large \bf  Radiation of charged particles\\
by charged black hole}
\end{center}

\begin{center}
I.B. Khriplovich\footnote{khriplovich@inp.nsk.su}
\end{center}
\begin{center}
Budker Institute of Nuclear Physics\\
630090 Novosibirsk, Russia,\\
and Novosibirsk University
\end{center}

\bigskip

\begin{abstract}
The probability of a charged particle production by the electric field 
of a charged black hole depends essentially on the particle energy.
This probability is found in the nonrelativistic and
ultrarelativistic limits. The range of values for the mass and 
charge of a black hole is indicated where the discussed mechanism of 
radiation is dominating over the Hawking one.
\end{abstract}

\vspace{8cm}

\end{titlepage}

{\bf 1.} The problem of particle production by the electric field
of a black hole has been discussed repeatedly~[1-6]. The probability of
this process was estimated in these papers using in some way or another
the result obtained previously~[7-9] for the case of an electric field 
constant all over the space. This approximation might look quite
natural with regard to sufficiently large black holes, for which the 
gravitational radius exceeds essentially the Compton wave length of
the particle $\lambda=1/m$. (We use in the present paper the units
with $\hbar=1,\;\;c=1$; the Newton gravitational constant
$k$ is written down explicitly.) 
However, in fact, as will be demonstrated below, the constant-field 
approximation, generally speaking, is inadequate to the present problem,
does not reflect a number of its essential peculiarities.
 
It is convenient to start the discussion just from the problem of 
particle creation by a constant electric field. Here and below we 
restrict to the consideration of the production of electrons and
positrons, first of all because the probability of emitting these
lightest charged particles is the maximum one. Besides, the picture 
of the Dirac sea allows one in the case of fermions to manage without
the second-quantization formalism, thus making the consideration most
transparent.
 
To calculate the main, exponential dependence of the effect, it is 
sufficient to restrict to a simple approach due to~\cite{sau} 
(see also the textbook~\cite{blp}). In the potential $- e E z$ of a 
constant electric field $E$ the usual Dirac gap (Fig. 1) tilts (see
Fig. 2). As a result, a particle which had a negative energy in the 
absence of the field, can now tunnel through the gap (see the dashed
line in Fig. 2) and go to infinity as a usual particle. The hole created 
in this way is nothing but antiparticle. An elementary calculation 
leads to the well-known result for the probability of particle 
creation:
\beq\label{con}
W \sim \exp\left(-{\pi m^2 \over e E}\right).
\eeq

This simple derivation explains clearly some important properties of 
the phenomenon. First of all, the action inside the barrier does not change
under a shift of the dashed line in Fig. 2 up or down. Just due to it
expression (\ref{con}) is independent of the energy of created 
particles. Then, for the external field to be considered as a constant
one, it should change weakly along the path inside the barrier. However, the
length of this path is not directly related to the Compton wave length
of the particle. In particular, for an arbitrary weak field the
path inside the barrier becomes arbitrary long. 
 
Thus, one may expect that the constant-field approximation is not, 
generally speaking, applicable to the problem of a charged black hole
radiation, and that the probability of particle production in this 
problem is strongly energy-dependent. The explicit form of this 
dependence will be found below. We restrict in the present work to the
case of a non-rotating black hole.
 
\bigskip
 
{\bf 2.} We start the solution of the problem with calculating the
action inside the barrier. The metric of a charged black hole is well-known:
\beq
ds^2=f dt^2-f^{-1}dr^2-r^2(d\theta^2+\sin ^2 \theta d\phi ^2),
\eeq
where
\beq
f=1-{2kM \over r}+{kQ^2 \over r^2},
\eeq
$M$ and  $Q$ being the mass and charge of the black hole, respectively.
The equation for a particle 4-momentum in these coordinates is 
\beq
f^{-1}\left(\ep - {eQ \over r}\right)^2-f p^2-{l^2 \over r^2}=m^2.
\eeq
Here $\ep$ and $p$ are the energy and radial momentum of the particle.
We assume that the particle charge $e$ is of the same sign as the 
charge of the hole $Q$, ascribing the charge $\;-e\;$ to the 
antiparticle.

Clearly, the action inside the barrier is minimum for the vanishing orbital
angular momentum $l$. It is rather evident therefore (and will be
demonstrated in the next section explicitly) that after the summation 
over $l$ just the $s$-state defines the exponential in the total
probability of the process. So, we restrict for the moment to the case
of a purely radial motion. The equation for the Dirac gap for $l=0$ is
\beq
\ep_{\pm}(r) = {eQ \over r} \pm m \sqrt{f}.
\eeq
It is presented in Fig. 3. It is known~\cite{der} that at the horizon
of a black hole, for $r=r_+=kM+\sqrt{k^2 M^2-k Q^2}\,$, the gap vanishes.
Then, with the increase of $r$ the lower boundary of the gap 
$\ep_{-}(r)$ decreases monotonically, tending asymptotically to $-m$.
The upper branch $\ep_{+}(r)$ at first, in general, increases, and then
decreases, tending asymptotically to $m$.
 
It is clear from Fig. 3 that those particles of the Dirac see whose 
coordinate $r$ exceeds the gravitational radius $r_+$ and whose energy
$\ep$ belongs to the interval $\ep_{-}(r) > \ep > m$, tunnel through
the gap to infinity. In other words, a black hole looses its charge due
to the discussed effect, by emitting particles with the same sign of
the charge $e$, as the sign of $Q$. Clearly, the phenomenon takes place 
only under the condition 
\beq
{eQ \over r_+} > m.
\eeq

For an extreme black hole, with $Q^2 = k M^2$, the Dirac gap looks 
somewhat different (see Fig. 4): when $Q^2$ tends to $k M^2$ the 
location of the maximum of the curve $\ep_{+}(r)$ tends to $r_+$, and
the value of the maximum tends to ${eQ/r_+}$. It is obvious however 
that the situation does not change qualitatively due to it. Thus, 
though an extreme black hole has zero Hawking temperature and, 
correspondingly, gives no thermal radiation, it still creates charged
particles due to the discussed effect.
  
In the general case $Q^2 \leq k M^2$ the doubled action inside the barrier
entering the exponential for the radiation probability is 
\[ 2S = 2\int_{r_1}^{r_2}dr\,|p(r,\ep)| \]
\beq 
= 2 \int_{r_1}^{r_2}{dr\,r \over r^2 - 2 k M r + k Q^2}
\sqrt{- p^2_0 r^2 + 2 (\ep e Q - k m^2 M) r - (e^2 - k m^2) Q^2}.
\eeq
Here $p_0=\sqrt{\ep^2-m^2}$ is the momentum of the emitted particle at
infinity, and the turning points $r_{1,2}$ are as usual the roots of 
the quadratic polynomial under the radical; we are interested in the 
energy interval $m\leq \ep \leq eQ/r_+$. Of course, the integral can be 
found explicitly, though it demands somewhat tedious calculations.
However, the result is sufficiently simple: 
\beq\label{ac}
2S = 2\pi\;{m^2 \over (\ep + p_0)\,p_0}\;[e\, Q - (\ep - p_0)\,k\,M].
\eeq
Certainly, this expression, as distinct from the exponent in
formula (\ref{con}), depends on the energy quite essentially.
 
Let us note that the action inside the barrier does not vanish even for the 
limiting value of the energy $\ep_m = eQ/r_+$. For a nonextreme black
hole it is clear already from Fig. 3. For an extreme black hole this 
fact is not as obvious. However, due to the singularity of $|p(r,\ep)|$,
the action inside the barrier is finite for $\ep = \ep_m = eQ/r_+\,$ for an 
extreme black hole as well. In this case the exponential factor in the 
probability is 
\beq
\exp\left(-\pi \,{\sqrt{k}m \over e}\, k m M \right).
\eeq
Due to the extreme smallness of the ratio 
\beq\label{re}
{\sqrt{k}m \over e} \sim 10^{-21}, 
\eeq
the exponent here is large only for a very heavy black hole, with a 
mass $M$ exceeding that of the Sun by more than 5 orders of magnitude.
And since the total probability, integrated over energy, is dominated
by the energy region $\ep \sim \ep_m$, the semiclassical approach is
applicable in the case of extreme black holes only for these very heavy
objects. Let us note also that for the particles emitted by an extreme
black hole, the typical values of the ratio $\ep / m$ are very large:
\[ {\ep \over m}\;\sim\; {\ep_m \over m}\; =\; {eQ \over k m M}
\; =\; {e \over \sqrt{k} m} \;\sim 10^{21}. \]
In other words, an extreme black hole in any case radiates highly
ultrarelativistic particles mainly. 

Let us come back to nonextreme holes. In the nonrelativistic limit, 
when $eQ/r_+ \rightarrow m$ and, correspondingly, the particle velocity
$v \rightarrow 0$, the exponential is of course very small:
\beq\label{nr}
\exp\left(- \,{2 \pi k m M \over v}\right).
\eeq
Therefore, we will consider mainly the opposite, ultrarelativistic limit
where the exponential is
\beq\label{ur}  
\exp\left(- \pi\, {m^2 \over \ep^2}\, e Q \right).
\eeq
Of course, here also the energies $\ep \sim \ep_m \sim eQ/kM$ are 
essential, so that the ultrarelativistic limit corresponds to the
condition
\beq\label{lim1}
eQ \gg kmM.  
\eeq
But then the semiclassical result (\ref{ur}) is applicable (i.e.,
the action inside the barrier is large) only under the condition 
\beq\label{lim2}
kmM \gg 1. 
\eeq
Let us note that this last condition means that the gravitational 
radius of the black hole ($r_+ \sim kM$) is much larger than the
Compton wave length of the electron $1/m$. In other words, the result
(\ref{ur}) refers to macroscopic black holes. Combining (\ref{lim1})
with (\ref{lim2}), we arrive at one more condition for the 
applicability of formula (\ref{ur}):
\beq\label{lim3}
eQ \gg 1.  
\eeq
We will come back to this relationship below.
 
Let us note that in~\cite{gib} the action inside the barrier was being
calculated 
under the same assumptions as formula (\ref{ur}). However, the answer
presented in~\cite{gib}, $2 S = \pi m^2 r_+^2/e Q$, is independent of
energy at all (and corresponds to formula (\ref{con}) which refers to
the case of a constant electric field). I do not understand how such an
answer could be obtained for the discussed integral in the general case 
$\ep \neq \ep_m$.

\bigskip

{\bf 3.} The obtained exponential is the probability that a particle
approaching the turning point $r_1$ (see Figs. 3, 4) from the left, will
tunnel through the potential barrier. One should recall that in the general
case the position of the turning point depends not only on the particle
energy $\ep$, but on its orbital angular momentum $l$ as well. The 
total number of particles with given $\ep$ and $l$, approaching a spherical 
surface of the radius $r_1$ in unit time, is equal to the product 
of the area of this surface
\beq\label{s}
S= 4 \pi\, r_1^2(\ep,\,l)
\eeq
times the current density of the particles
\beq\label{cud}
j^r(\ep,\,l) = \;{\rho \over \sqrt{g_{00}}}\;{dr \over dt}
\eeq
(see, e.g.,~\cite{ll}, \S 90). The particle velocity is as usual
\beq
v^r = \;{dr \over dt}\;=\;{\partial \ep \over \partial p}
\eeq 
(the subscript $r$ of the radial momentum $p$ is again omitted).
To obtain an explicit expression  for the particle density $\rho$, we
will use the semiclassical approximation (the conditions of its 
applicability for the region $r_+\leq r \leq r_1$ will be discussed
later). Let us note that the volume element of the phase space 
\beq\label{phs}
2\;{dp_x dp_y dp_z dx dy dz \over (2\pi)^3} 
\eeq
is a scalar. (The factor 2 here is due as usual to 
two possible orientations of the electron spin.) On the other hand,
the number of particles in the elementary cell $dx dy dz$ equals 
(see
 \cite{ll}, \S 90)
\beq\label{pan}
\rho \sqrt{\gamma} dx dy dz, 
\eeq
where $\gamma$ is the determinant of the space metric tensor. Since all
the states of the Dirac sea are occupied, we obtain by comparing
formulae (\ref{phs}) and (\ref{pan}) that the following expression
\[ {\rho \over \sqrt{g_{00}}}\;=\;
{2 \over \sqrt{g_{00} \gamma}}\; \sum \;{dp_x dp_y dp_z \over (2\pi)^3}\;=\;
{2 \over \sqrt{-g}}\;\sum \;{dp_x dp_y dp_z \over (2\pi)^3} \]
should be plugged in formula (\ref{cud}) for the current density
(the summation here and below is performed with fixed $\ep$ and $l$,
see (\ref{cud})).
In our case the determinant $g$ of the four-dimensional metric tensor
does not differ from the flat one, so that the radial current density
of the particles of the Dirac sea is
\beq\label{cud1}
j^r(\ep,\,l) = 2 \sum \;{d^3 p \over (2\pi)^3}\;
{\partial \ep \over \partial p}.
\eeq
The summation in the right-hand-side reduces in fact to the 
multiplication by the number $2l+1$ of possible projections of the 
orbital angular momentum ${\bf l}$ onto the $z$ axis and to the
integration over the azimuth angle of the vector ${\bf l}$, which gives
$2\pi$. With the account for the identity
\[ {\partial \ep \over \partial p_r}\,dp_r\,=\,d\ep, \] 
we obtain in the result
\beq\label{cud2}
j^r(\ep,\,l) = 2\;{2\pi (2l+1) \over (2\pi)^3 r_1^2(\ep,\,l)}.
\eeq
Finally, the pre-exponential factor in the probability, differential in
energy and orbital angular momentum, is
\beq
{2 (2l+1) \over \pi}\,.
\eeq
Correspondingly, the number of particles emitted per unit time is
\beq\label{n/t}
{dN \over dt}\;=\;{2 \over \pi}\;\int d\ep \sum_l (2l+1) 
\exp [-2S(\ep, l)].
\eeq 

In the most interesting, ultrarelativistic case $dN/dt$ can be 
calculated explicitly. Let us consider the expression for the momentum
in the region inside the barrier for $l \neq 0$
\beq\label{p}
|p(\ep, l, r)|= f^{-1}\sqrt{\left(m^2 + {l^2 \over r^2}\right)f
- \left(\ep - {eQ \over r}\right)^2}.
\eeq 
The main contribution to the integral over energies in formula 
(\ref{n/t}) is given by the region $\ep \rightarrow \ep_m$. In this
region the functions $f(r)$ and $\ep - eQ / r$, entering expression
(\ref{p}), are small and change rapidly. As to the quantity 
\beq\label{mu}
\mu^2(r,l) = m^2 + {l^2 \over r^2},
\eeq
one can substitute in it for $r$ its average value, which lies between
the turning points $r_1$ and $r_2$. Obviously, in the discussed limit
$\ep \rightarrow \ep_m$ the near turning point coincides with the 
horizon radius, $r_1= r_+$. And the expression for the distant turning 
point is in this limit
\beq
r_2=\,r_+\left[1 +\;{2\mu^2 \over \ep_m^2 - \mu^2}\;
{\sqrt{k^2 M^2 - k Q^2} \over r_+}\;\right].
\eeq 
Assuming that for estimates one can put in formula (\ref{mu}) 
$r \sim r_+$, one can easily show that the correction to 1 in the
square bracket is bounded by the ratio $l^2/(eQ)^2$. Assuming that this 
ratio is small (we will see below that this assumption is 
self-consistent), we arrive at the conclusion that $r_2\approx r_+$, 
and hence $\mu^2$ can be considered independent of $r$: 
$\mu^2(r,l) = m^2 + l^2 / r_+^2$. As a result, we obtain
\beq\label{}
2S(\ep, l)\approx \pi eQ \left(\;{m^2 \over \ep^2}\;+\; 
{l^2 \over r_+^2 \ep^2}\;\right).
\eeq
Now we find easily 
\beq\label{n/t1}
{dN \over dt}\;=\,m\left({eQ \over \pi m r_+}\right)^3 
\exp \left(-\;{\pi m^2 r_+^2 \over eQ}\right).
\eeq  
Let us note that the range of orbital angular momenta, contributing to
the total probability (\ref{n/t1}), is effectively bounded by the
condition $l^2 \leq eQ$. Since $eQ \gg 1$, this condition allows one
to change from the summation over $l$ in formula (\ref{n/t}) to the
integration. On the other hand, this condition justifies the used
approximation $\mu^2(r,l) = m^2 + l^2 / r_+^2$.

However, up to now we have not considered one more condition necessary
for the derivation of formula (\ref{n/t1}). We mean the applicability
of the semiclassical approximation to the left of the barrier, for
$r_+ \leq r \leq r_1$. This condition has the usual form
\beq\label{wkb}
{d \over dr}\,{1 \over p(r)} < 1. 
\eeq
In other words, the minimum size of the initial 
wave packet should not exceed
the distance from the horizon to the turning point. Using the estimate
\[ p(r) \sim \;{r_+\,(eQ - \ep r_+) \over (r-r_+)(r-r_-)} \] 
for the momentum in the most essential region, one can check that for
an extreme black hole the condition (\ref{wkb}) is valid due to the 
bound $eQ \gg 1$. In a non-extreme case, for $r_+\;-r_-\;\sim r_+$, the
situation is different: the condition (\ref{wkb}) reduces to 
\beq\label{wkb1}
\ep < {eQ - 1 \over r_+}\; \sim\;{eQ \over r_+}.
\eeq
Thus, for a non-extreme black hole in the most essential region
$\ep \rightarrow \ep_m$ the condition of the semiclassical 
approximation is not valid. Nevertheless, the semiclassical result 
(\ref{n/t}) remains true qualitatively, up to a numerical factor in the
pre-exponential. 

In concluding this section few words on the radiation of light
charged black holes, for which $kmM <1$, i.e., for which the
gravitational radius is less than the Compton wave length of the
electron. In this case the first part, 
\[ \ep < {eQ - 1 \over r_+}, \]
of inequality (\ref{wkb1}), which guarantees the localization of the
initial wave packet in the region of a strong field, means in 
particular that  
\beq\label{cr}
eQ\,=\,Z\alpha > 1  
\eeq
(we have introduced here $Z=Q/e$). It is well-known (see, 
e.g.,~\cite{zep,mig}) that the vacuum for a point-like charge with
$Z\alpha > 1$ is unstable, so that such an object looses its charge
by emitting charged particles. It is quite natural that for a black 
hole whose gravitational radius is smaller than the Compton wave length 
of the electron, the condition of emitting a charge is the same as in 
the pure quantum electrodynamics. (Let us note that the unity in all
these conditions should not be taken too literally: even in quantum
electrodynamics, where the instability condition for the vacuum of
particles of spin 1/2 is for a point-like nucleus just $Z\alpha > 1$, 
for a finite-size nucleus it changes~\cite{zep,mig} to 
$Z\alpha > 1.24$. On the other hand, for the vacuum of scalar particles
in the field of a point-like nucleus the instability condition 
is~\cite{som,bet}: $Z\alpha > 1/2$.) As has been mentioned 
already, for a light black hole, with $kmM <1$, the discussed condition
$eQ > 1$ leads to a small action inside the barrier and to the 
inapplicability of the semiclassical approximation used in the present
article. The problem of the radiation of a charged black hole with 
$kmM <1$ was investigated numerically in~\cite{pag}.

\bigskip

{\bf 4.} The exponential
\[ \exp \left(-\;{\pi m^2 r_+^2 \over eQ}\right) \]
in our formula (\ref{n/t1}) coincides with the expression arising from
formula (\ref{con}), which refers to a constant electric field $E$,
if one plugs in for this field its value $Q/r_+^2$ at the black hole
horizon. As has been mentioned already, an approach based on formulae
for a constant electric field was used previously in Refs.~[1-6]. Thus,
our result for the main, exponential dependence of the probability
integrated over energies, coincides with the corresponding result of 
these papers. Moreover, our final formula (\ref{n/t}) agrees with the
corresponding result of Ref.~\cite{nos} up to an overall factor 1/2. 
(This difference is of no interest by itself: as has been noted above,
for a non-extreme black hole the semiclassical approximation cannot
guarantee at all an exact value of the overall numerical factor.) 

Nevertheless, we believe that the analysis of the phenomenon performed 
in the present work, which demonstrates its essential distinctions from
the particle production by a constant external field, is useful. First
of all, it follows from this analysis that the probability of the 
particle production by a charged black hole has absolutely nontrivial 
energy spectrum. Then, in no way are real particles produced by a 
charged black hole all over the whole space: for a given energy $\ep$
they are radiated by a spherical surface of the radius $r_2(\ep)$, 
this surface being close to the horizon for the maximum energy. (It 
follows from this,
for instance, that the derivation of the mentioned result of 
Ref.~\cite{nos} for $dN/dt$ has no physical grounds: this derivation 
reduces to plugging
$E=Q/r^2$ into the well-known Schwinger formula~\cite{sch}, obtained
for a constant field, with subsequent integrating all over the space
outside the horizon.)

Let us compare now the radiation intensity $I$ due to the effect 
discussed, with the intensity $I_H$ of the Hawking thermal radiation.
Introducing additional weight $\ep$ in the integrand of formula 
(\ref{n/t}), we obtain
\beq\label{I}
I =\,\pi m^2\left({eQ \over \pi m r_+}\right)^4 
\exp \left(-\;{\pi m^2 r_+^2 \over eQ}\right).
\eeq   
As to the Hawking intensity, the simplest way to estimate it, is to
use dimensional arguments, just to divide the Hawking temperature 
\[ T_H =\;{1 \over 4\pi r_+} \]
by a typical classical time of the problem $r_+$ (in our units $c=1$).
Thus,
\beq\label{I_H}
I_H \sim \,{1 \over 4\pi r_+^2}.
\eeq   
More accurate answer for $I_H$ differs from this estimate by a small
numerical factor $\sim 2\cdot 10^{-2}$, but for qualitative estimates
one can neglect this distinction. The intensities (\ref{I}) and 
(\ref{I_H}) get equal for
\beq
eQ \sim \;{\pi \over 6}\;{(m r_+)^2 \over \ln(m r_+)}\;\sim\;
{\pi \over 6}\;{(k m M)^2 \over \ln(k m M)}.
\eeq
(One cannot agree with the condition $eQ \sim 1/(4\pi)$ for the equality 
of these intensities, derived in Ref.~\cite{nos} from the comparison 
of $\ep_m =eQ/r_+$ with $T_H = 1 /(4\pi r_+)\,$.)

Let us consider in conclusion the change of the horizon surface of a
black hole, and hence of its entropy, due to the discussed non-thermal
radiation. To this end, it is convenient to introduce, following
Ref.~\cite{chr}, the so-called irreducible mass $M_0$ of a black hole:
\beq
2 M_0 = M + \sqrt{M^2-Q^2};
\eeq
here and below we put $k=1$. This relationship can be conveniently
rewritten also as       
\beq\label{irr}
M = M_0 + {Q^2 \over 4M_0}.
\eeq  
Obviously, $r_+\;=2M_0$, so that the horizon surface and the black 
hole entropy are proportional to $M_0^2$.

When a charged particle is emitted, the charge of a black hole changes
by $\Delta Q = -e$, and its mass by $\Delta M = - eQ/r_+\,+\xi$, where
$\xi$ is the deviation of the particle energy from the maximum one.
Using the relationship (\ref{irr}), one can easily see that as a result
of the radiation, the irreducible mass $M_0$, and 
hence the horizon surface and entropy of a non-extreme black hole do 
not change if the particle energy is the maximum one $eQ/r_+$. In other
words, such a process, which is the most probable one, is adiabatic. For
$\xi > 0$, the irreducible mass, horizon surface, and entropy increase.

As usual, an extreme black hole, with $M=Q=2M_0$, is a special case.
Here for the maximum energy of an emitted particle $\ep_m = e$, we have
$\Delta M = \Delta Q = -e$, so that the black hole remains extreme
after the radiation. In this case $\Delta M_0 = - e/2$, the irreducible 
mass and the horizon surface decrease. In a more general case, 
$\Delta M = - e + \xi$, the irreducible mass changes as follows: 
\beq
\Delta M_0 = -\,{e-\xi \over 2}\;+
\sqrt{\left(M_0-\,{e \over 2}\,+\,{\xi \over 4}\right)\xi}.
\eeq
Clearly, in the case of an extreme black hole of a large mass, 
already for a small deviation $\xi$ of the emitted energy from the 
maximum one, the square root is dominating in this expression, so that 
the horizon surface increases. 

\bigskip
\bigskip
\bigskip

I am grateful to I.V. Kolokolov, A.I. Milstein, V.V. Sokolov, and
O.V. Zhirov for the interest to the work and useful comments. The work
was supported by the Russian Foundation for Basic Research through
Grant No. 98-02-17797, and by the Federal Program Integration-1998
through Project No. 274.

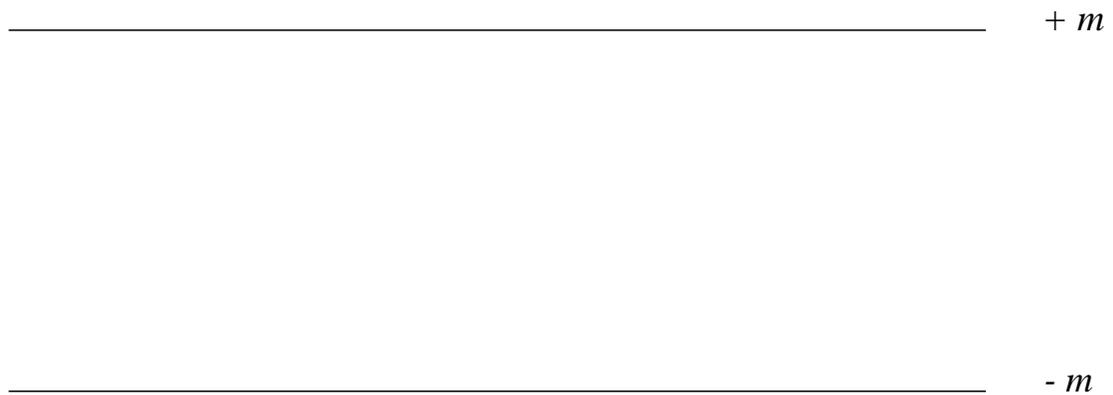
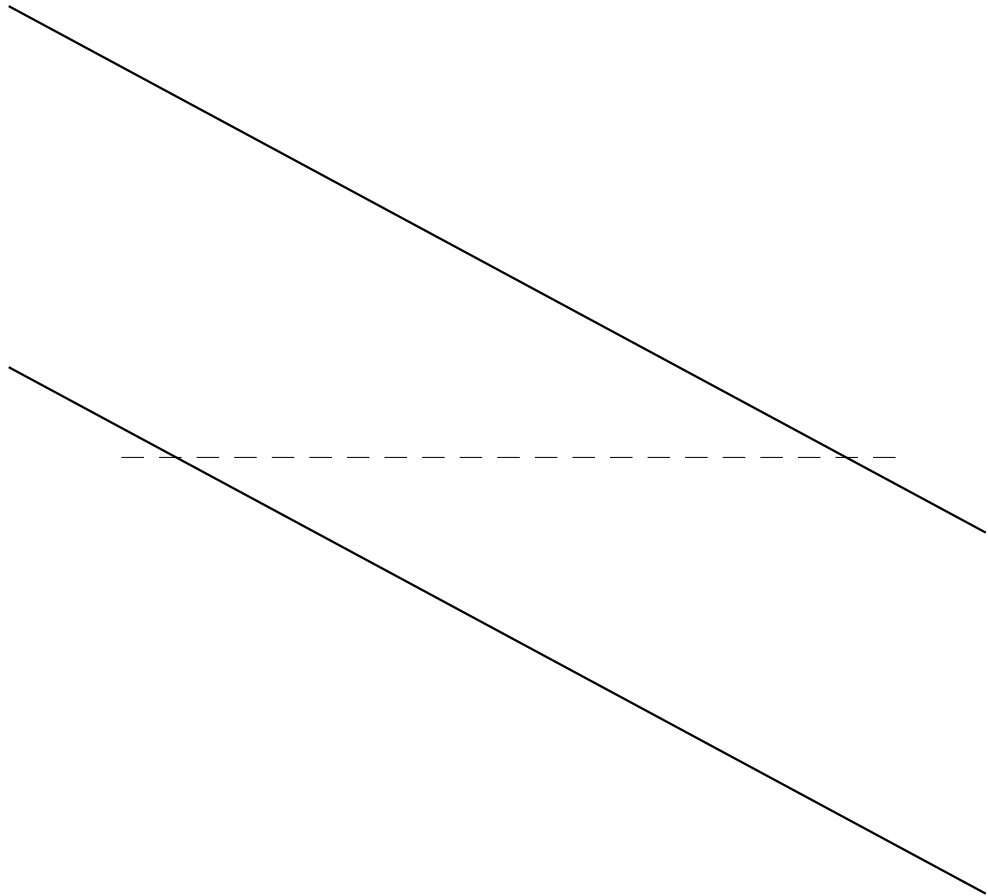
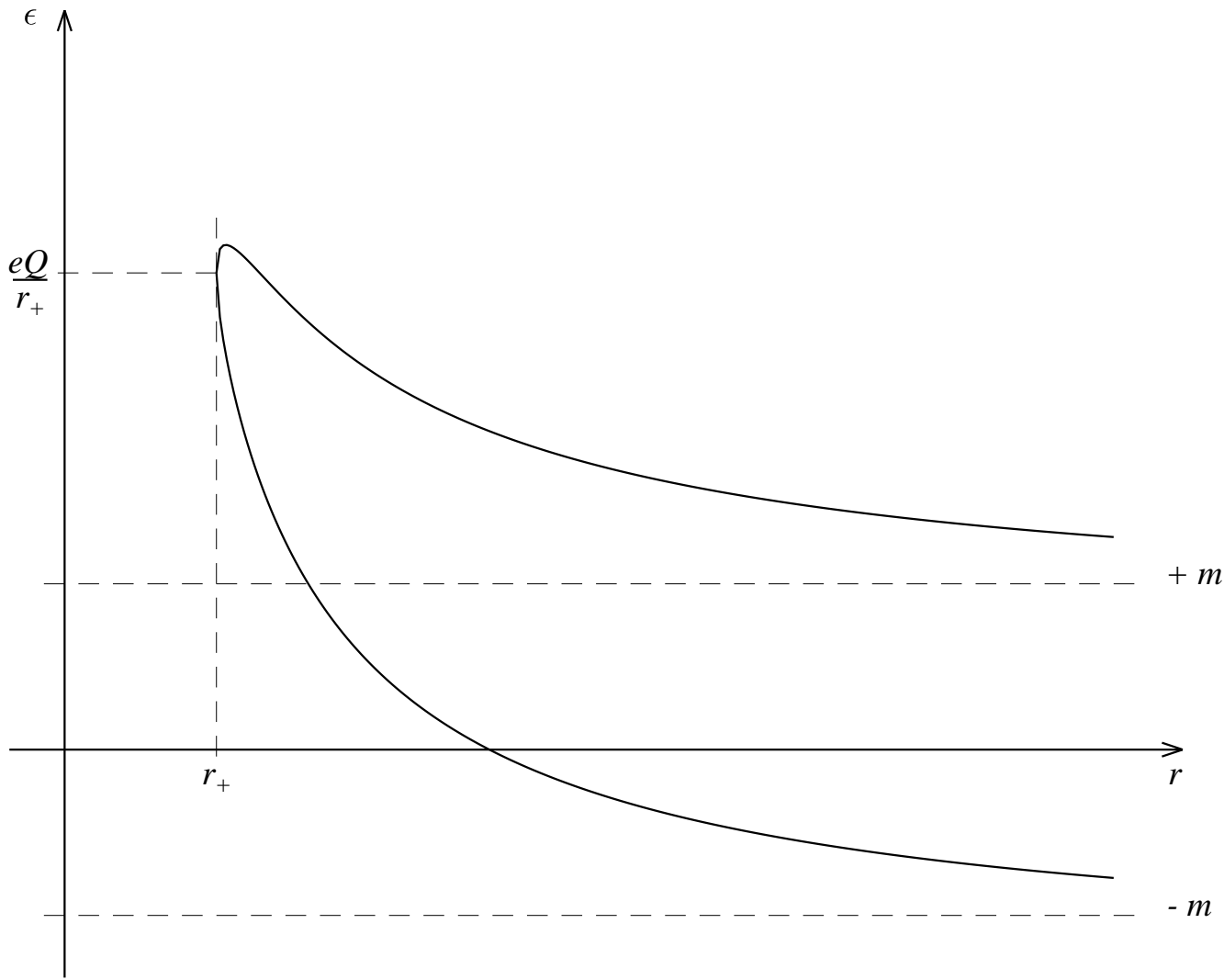
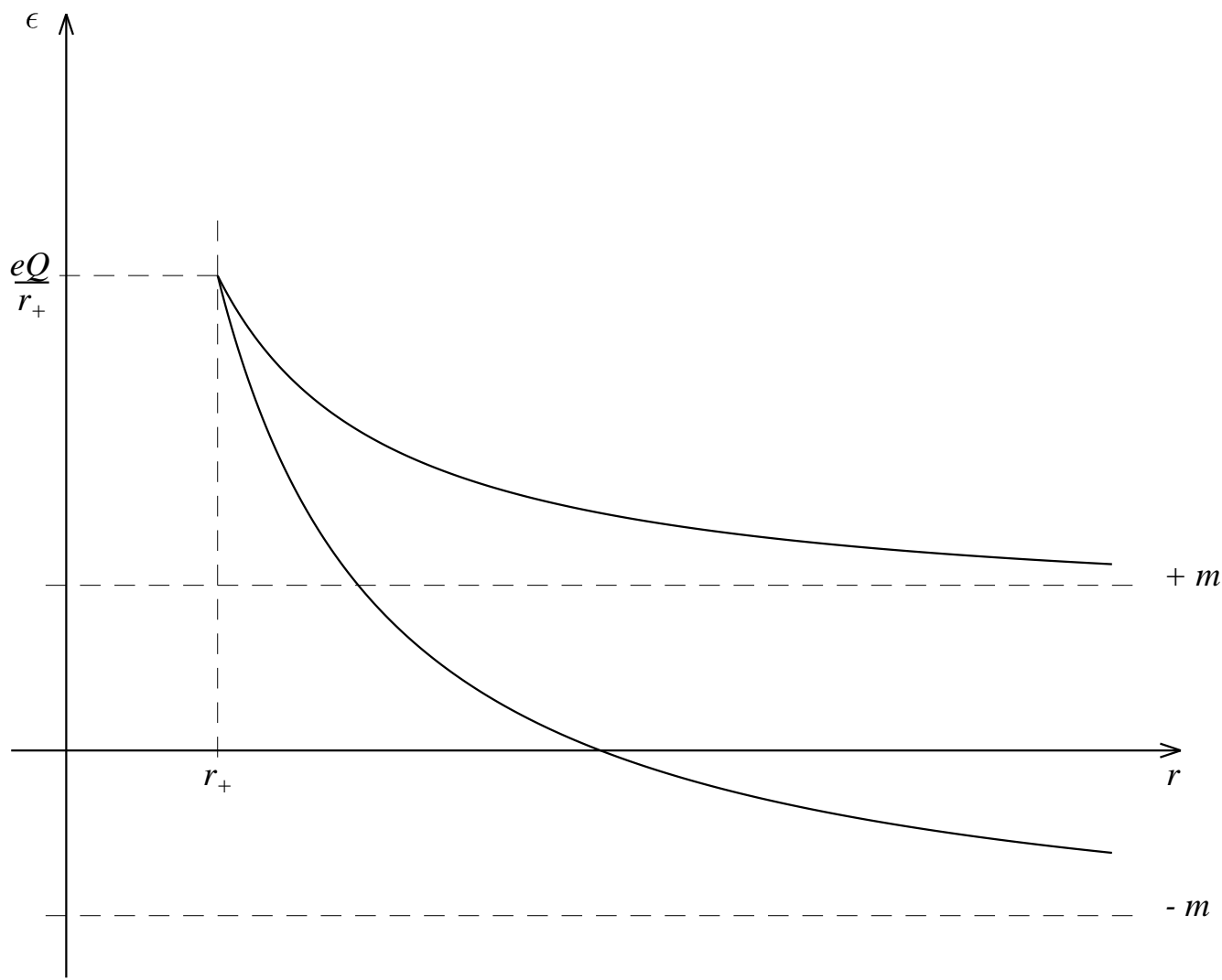
\end{document}